\documentstyle[aps,preprint,epsfig]{revtex}
\begin{document}
%\preprint{NSF-ITP-97-114}
\draft

\title{\Large{Probing the ladder spectrum arising from motion in a
 2-D lattice driven by an in-plane constant force}}

\author{W. Yan$^{1,2}$, F. Claro$^{1}$, Z.Y. Zeng$^{1}$, Y.L. Zhao$^{2}$ and J.Q. Liang$^{2}$}
\address{$^1$Facultad de F\'isica, Pontificia Universidad de Cat\'olica de Chile, Casilla 306, Santiago 22, Chile}
\address{$^2$Institute of Theoretical Physics, Shanxi University,
Taiyuan, 030006, People's Republic of China}

%\date{\empty}
\maketitle

\begin{abstract}
The coherent interband dynamics of optically excited
two-dimensional lateral surface superlattices driven by an
in-plane static electric field has been investigated. The linear
absorption, the spectrally-resolved pump-probe four-wave mixing
signals and spatial coherent wavepacket evolution in the
time-domain are obtained. When the rational condition
$E_x/E_y=p/q$, with $p,q$ prime to each other, is fullfilled, it
is found that $p$ peaks appear within the frequency interval
$\omega_{Bx}=eE_xa/\hbar$ in both linear absorption and degenerate
four-wave mixing signals. The coherent time evolution of the
electron-hole pair wavepacket is that of a breathing mode, with
the period $2\pi p/\omega_{Bx}$, These findings are consistent
with the recent spectral results (Phys. Rev. Lett. {\bf 86},
3116), hence providing a method for probing the coherent dynamics
of quantum particles in 2D lattices.

\end{abstract}


%\widetext
%\narrowtext

\section{Introduction}

Recently, there have been much investigation on the electronic and
transport properties of semiconductor superlattices driven by the
external static and/or time-dependent electric
fields\cite{mendez,meierl94,meierl95,axt,whittaker,keay,unter,dunlap,jauho,loser,zhao,dignamprb99,liuprb,hader}.
It is foreseen that these explorations will uncover rich physics
and even find potential applications in electro-optic devices. In
this context, a lot of interesting theoretical and experimental
findings have been reported. Among these, we mention Bloch
oscillations\cite{mendez}, Fano resonances\cite{whittaker},
absolute negative conductivity\cite{keay}, inverse Bloch
oscillators\cite{unter}, dynamic localization\cite{keay,dunlap},
dynamic Franz-Keldysh effect\cite{jauho}, self-induced Shapiro
effect\cite{loser},
 dynamic fractional Wannier-Stark ladders\cite{zhao}. Particularly,
 with the development of atom optics, physicists began to link
 solid state physics and optics together, both theoretically
 and experimentally\cite{raizenpt}. The recent {\it clean} demonstration of Wannier-Stark
 ladders\cite{wilkinson} and Bloch oscillations\cite{dahan} in the atom optics are typical
 examples.

Several months ago, Gl\"uck et al calculated numerically the
ground Wannier-Bloch band of a quantum particle moving in a
lattice under an in-plane electric field\cite{gluck}, The lattice
may be an array of quantum dots for electrons, or an interference
pattern of light waves for cold atoms\cite{raizenpt}. They found
that the energy spectrum resembles the Hofstader
butterfly\cite{hofstader} in a magnetic system, and the 1-D
superlattice driven by dc-ac electric fields\cite{zhao}. They
showed that for a rational static force ${\bf F}$ of components in
the ratio $F_x/F_y=p/q$, with $p,q$ coprime integers, there are
$s=p^2+q^2$
 identical subbands separated by the energy interval
 $\delta\equiv|{\bf F}|a/\sqrt{s}$, where $a$ is the lattice constant. Since the
 lattice period in the direction of the field is $\sqrt{s}a$, this
  result predicts a splitting of the 1-D Stark period into $s$
  equally spaced subbands. The question arises as to whether such a
  spectrum is measurable.

 In this paper we report our finding that
this striking splitting in the spectrum can be identified in
optically excited two-dimensional lateral surface superlattices
driven by an in-plane static electric field ${\bf E}$. Such 2-D
 superlattices can be prepared by embedding a square array of
 GaAs  cylinders in a very thin Ga$_{x}$Al$_{1-x}$As
epilayer\cite{reich}. By numerically solving the generalized
semiconductor Bloch equations, this spectrum is identified in both
the linear absorption signal and spectrally resolved pump-probe
degenerate four-wave mixing signals. Also, we link the
breathing-mode period of spatial coherent wavepackets oscillations
in the time domain in the system to the splitting of the spectrum.

\section{Model and numerical results}
 We consider an electron system moving in a 2-D lattice and an
 external in-plane electric field ${\bf E}$ with its component in
 the ratio $E_x/E_y=p/q$, where $p$ and $q$ are coprime numbers.
 The energy dispersion is treated in a two-band model, the valence and
  conduction bands. We discuss the dynamics in the context of
 the following semiconductor Bloch
 equations\cite{meierl94,meierl95,axt,lindberg}

\begin{eqnarray}
\left(\frac{\partial}{\partial t}+{\frac{e}{\hbar}\bf
E}(t)\cdot{\bigtriangledown_{{\bf k}}}\right)P_{\bf k}(t)
&=&-\frac{i}{\hbar}[e_{e,{\bf k}}+e_{h,{\bf
k}}-i\Gamma_L]P_{\bf k}(t) \nonumber \\
&-& \frac{i}{\hbar}[n_{e,{\bf k}}+n_{h,{\bf k}}-1] \omega_{R,{\bf
k}}
\end{eqnarray}

\begin{eqnarray}
\left(\frac{\partial}{\partial t}\pm\frac{e}{\hbar}{\bf
E}(t)\cdot{\bigtriangledown}_{\bf{ k}}\right)n_{e(h)\bf k}(t)
&=&-2{\rm {Im}}[\omega_{R,\bf k}P^*_{\bf k}] \nonumber \\
&-&\Gamma_T{n_{e,(h){\bf k}}(t)},
\end{eqnarray}

\noindent where $P_{\bf k}(t)$ is the interband polarization and
$n_{e(h),\bf k}(t)$ the electron (hole) population density in the
conduction (valence) band. The quantities $e_{i,{\bf
k}}=\epsilon_{i,{\bf k}}-\sum_{\bf q}V_{|{\bf k}-{\bf
q}|}n_{i,{\bf q}}, ~(i=e,h)$ are the renormalized electron and
hole energies due to the Coulomb interaction. Also,
 $\omega_{R,\bf k}=(d_{cv}{\cal E}+\sum_{{\bf k}}V_{|{\bf k}-{\bf q}|}
P_{\bf q})/\hbar$ are the renormalized Rabi frequencies, with
$d_{cv}$ the dipole moment and
 ${\cal E}(t)$, the Gaussian laser pulses.
 The relaxation time approximation has been assumed, with $\Gamma_L$ and
$\Gamma_T$ being the longitudinal and transverse relaxation rates,
respectively.

 The spectrally-resolved absorption, pump-probe nonlinear
four-wave mixing signals and the real-time coherent evolution of
the wavepackets in real-space have been calculated numerically by
solving the coupled Eqs.(1) and (2). The energy bands of 2-D
lateral surface superlattices of square geometry can be described
by the tight-binding form\cite{reich,silin,davies}
$\epsilon_i({\bf
k})=\frac{\Delta_i}{2}(\cos(k_xa)+\cos(k_ya))~(i=e,h)$, with
$\Delta_c(\Delta_v)$ being the combined miniband width of the
conduction (valence) band in the $x$ and $y$ directions. For
simplicity, an on-site Coulomb interaction has been adopted, which
 describes the first-order Born scattering among the carriers.
 Although the excitonic interaction is simplified to be on-site
   in this work, it is expected that the inclusion of more
  realistic models will not change our findings
  qualitatively.
 This kind of approximation has been used with success by
 many groups\cite{meierl94,meierl95,axt,liuprb,yanjpcmprb}.

\textit{Linear absorption}: In the linear response regime, one can
completely neglect Eq.(2), and set the electron (hole) density
$n_{c(v)}({\bf k},t)$ to vanish in Eq. (1). The absorption
$\alpha(\omega)$ is proportional to the imaginary part of the
first-order susceptibility
$\chi^{(1)}(\omega)=\rm{Im}[P^{(1)}(\omega)/E(\omega)]$, where
 $P^{(1)}(\omega)$ is the Fourier transform of the quantity $\sum_{\bf k}P^{(1)}({\bf k},t)$.
 The integro-differential equations can be solved numerically in
 the accelerated basis ${\bf k}-\frac{e}{\hbar}\int_0^t{\bf E}(\tau)d\tau$. Parameter
  values used in the calculations
  are $\Delta_c=\Delta_v=20$ meV, $V=10$ meV, and
 $\Gamma_L=\Gamma_T=0.5$ THz. The full width at half maximum for
 the strength of the Gaussian laser pulse is assumed to be $59$
  femtoseconds.
 In the simulation the Bloch frequency $\omega_{Bx}=eE_xa/\hbar$ of the
  static electric field in the $x$ direction remains unchanged and taken to be $3\pi$
   THz, while the Bloch frequency $\omega_{By}=eE_y/\hbar$ in the $y$
 direction takes three different values: $1/3~\omega_{Bx}$, $2/3~
 \omega_{Bx}$, and $4/3~\omega_{Bx}$. Distinct peaks are clearly identified
 in the three panels, where the ratio $E_x/E_y$ is indicated in the right-up corner.
  Roughly three peaks appear within every interval
  $\omega_{Bx}$ in the three panels, although the Bloch frequencies
 in the $y$ direction differ by a factor of four in the top and
  bottom panels. This property is the direct reflection
  of the striking spectrum discussed by Gl\"uck et al
  \cite{gluck}. Since $\omega_{Bx}=eE_xa/\hbar\equiv (p|e{\bf E}|/\sqrt{s})(a/\hbar)$,
  and the characteristic energy splitting $\delta=|{e\bf E}|a/\hbar\sqrt{s}$, then
  $\delta/\omega_{Bx}=1/p$, where for Fig.1, $p=3$ in
   all three panels. The interesting and special case of
   $E_x:E_y=3:3=1:1$ is shown in the Fig.2, where peaks appear
   only once in the fixed interval $\omega_{Bx}$. It is clear from
   the comparison of Fig.1 and Fig.2 that the number of peaks appearing in the fixed
    frequency interval $\omega_{Bx}$ is solely determined by the integer $p$,
    instead of the absolute value of $E_x$ and $E_y$.
   It should be noted that the absorption peaks obtained in the present model
 are much more stable that those in dc-ac field case\cite{yanjpcm01}.
 The latter are easily blurred by the dephasing and Coulomb interaction-mediated
 Fano interferences\cite{whittaker}.

\textit{Four-wave mixing signals}: Four-wave mixing experiments
are usually applied to observe the coherent signals from optically
excited semiconductor samples by taking advantage of the fact that
this technique can substantially avoid the inhomogeneous
broadening\cite{meierl94}. In order to mimic the experimental
realization of probing the spectrum,
 the two laser pulses: the strong pump pulse propagating in the ${\bf k}_1$ direction, and
 the weak probe pulse propagating in the direction ${\bf k}_2$, are
 arranged to be delayed a time $\tau_p$ relative to each other,
\begin{eqnarray}
{\cal E}(t)&=&{\cal E}_1(t)\exp[i({\bf k}_1\cdot{\bf r}-\omega
t)]\nonumber \\ &+&{\cal E}_2(t-\tau_p)\exp[i({\bf k}_2\cdot{\bf
r}-\omega (t-\tau_p))]~.
\end{eqnarray}
One can use the method of Lindberg et al\cite{lindbergpra}, or
that by Banyai et al\cite{banyai}, to extract the nonlinear
pump-probe four-wave mixing signals propagating in the direction
$2{\bf k}_2-{\bf k}_1$. The spectrally-resolved pump-probe
degenerate four-wave mixing squared signals are shown in the four
panels of Fig.3 for different values of $\tau_p$. A Gaussian
profile for the pump and probe laser fields ${\cal E}_1(t)$ and
${\cal E}_2(t)$ is assumed, and the other parameters used are the
same as those in Fig.1.
 In the figure, we only show the case of
$\omega_{By}/\omega_{Bx}=2/3$. The other sample cases
$\omega_{By}/\omega_{Bx}=1/3,4/3$ have similar shape, and are
omitted for saving space. Inspection of the four panels clearly
show that the peaks also appear three times in the frequency
interval $\omega_{Bx}$, although a different time delay $\tau_p$
is employed in different panels. This demonstrates that there
exists the energy interval/ladders in the system whose energy
spacing is identical and equal to $\omega_{Bx}/3$. This is a
direct manifestation of the split stable resonance spectrum
 found in 2-D lattices under an external in plane uniform field.

\textit{Coherent wavepacket oscillations}: As is well known, the
time-domain wavepacket Bloch oscillations inferred from the
semiclassical approach are a direct proof of the
 frequency-domain resonance states of Wannier-Stark ladders\cite{mendez}.
 While in Fig.1-3 the signals are all from the frequency domains,
 what about the evolution of the coherent wavepacket in the time and
  spatial domain? In the following, we will show the coherent
 spatial wavepacket oscillatory behavior in the realm-time domain\cite{note1}.

  According to the definition of the electron-hole
  wavepackets\cite{hughes}:
 $P({\bf r},t)=|\sum_{\bf k}\exp(-i{\bf k}\cdot{\bf r})P_{\bf
k}(t)|^2$, where
 ${\bf r}$ is the relative position vector of the electron-hole
 pair.
 The coherent spatial wavepackets have been
  shown sequentially in Fig.4, with the corresponding time
  displayed in the upper-right corners. In the simulation we used
  the following parameters: $\omega_{Bx}=6\pi$ THz, $E_y/E_x=2/3$, while the other
  parameters used are the same as those in Fig.1. The contour of the wavepacket in
 the earliest time ($t=0.05$ ps)
  looks like a small circular dot, which can be regarded as the profile of the initial
  excitation of the electron-hole pair. With the development of
  time ($t=0.20$ ps), due to the driving electric field,
  the wavepacket is pulled apart and distributed along both the $x$ and $y$ directions
  in a anisotropic manner. When time reaches $0.5$ ps, the wavepacket
  distribution along the $y$ direction is reduced. This is
  expected because the Bloch oscillations period along the $y$
  direction is $T_{By}=2\pi/\omega_{Bx}=0.5$ ps. For the same reason, the wavepacket
  distribution experiences narrowing along the direction $x$ at time
  $t=0.67~ps(\approx2T_{Bx})$. At time $t=0.8$ ps, the wavepacket again inflates
  in both $x$ and $y$ directions. The most striking phenomena perhaps lies in
  the last panel, where the wavepacket has shrunk into a small
  constellate at time $t=1$ ps. Since $T=2\pi\hbar/\delta=2\pi
  p/\omega_{Bx}$, for the frequency chosen one obtains $T=1$ ps,
  which equals three times $T_{Bx}$, and twice $T_{By}$.
 This indicates that the energy splitting $\delta$ found in
 Ref.\onlinecite{gluck} is the direct consequence of the wavepacket Bloch oscillations
 with period $T$ of the electron-hole pair, although no hole motions are involved in Ref.\onlinecite{gluck}.
The existence of this period is thus a direct manifestation of
 \textit{commensurate} synchronous Bloch oscillations in both $x$
 and $y$ directions. It should be mentioned in
 passing that in the absence of excitonic interaction, the wavepacket will
 undergo the breathing mode in a precise time period $T$. The
 presence of the interaction changes only slightly this period.

\section{Concluding Remarks} The linear absorption and nonlinear
pump-probe spectrally resolved four-wave mixing signals have been
 calculated in optically excited 2-D lateral surface
 superlattices driven by an in-plane static electric field.
 The ratio of $E_{x}$ to $E_{y}$ is chosen to
 be the rational $p/q$, making the peaks appear $p$ times
  in the frequency interval $\omega_{Bx}$, both in linear
  absorption and the four-wave mixing signal. This finding provides a way
   to test experimentally the spectrum arising in the dynamics of particles moving in a
    2D lattice and a constant force\cite{gluck}.
  Also the coherent wavepacket oscillates with a period corresponding
  exactly to that associated with the energy interval $\delta$,
  showing
  that the energy interval is a consequence of the commensurate
  \textit{synchronous} Bloch oscillations in both the $x$ and $y$ directions.

Theoretically, in the absence of all kinds of
dephasing\cite{gluck}, the spectrum has the property of extreme
sensitivity to fluctuations in the direction of the field. For
instance, if $E_y/E_x=q/p$, then an arbitrarily small rotation
yields a ratio $q^\prime/p^\prime$, with
$(q^\prime,p^\prime)\rightarrow\infty$.
 For instance, if $q/p=1/3$, a tiny change in the field produces the fraction
 $1001/3000$, while an even smaller change gives rise to the ratio
 $1000001/3000001$,  thus modifying wildly the splitting size as determined
 by the denominator fraction.
 The presence
 of various kinds of dephasing mechanisms blur this fine structure, making the
 identification of the peak in the fixed interval $\omega_{Bx}$ impossible, when
 $p$ becomes large\cite{note2}.
  Experimentally, the phenomena predicted here can be observed in a photoexcited 2D
 quantum dot array driven by an in-plane electric field, with the electric field
 strength tuned to be a rational. It is expected that \textit{clean} observation of
 these ladders can be also realized in ultracold atoms dwelling in
 the two perpendicular optical waves, accelerated accordingly.

\begin{center}
{\bf ACKNOWLEDGMENT}
\end{center}

This work was supported in part by C\'atedra Presidencial en
Ciencias (F.C.) and FONDECYT, Grants 1990425 and 3980014; W.Y. is
 supported in part by the Youth Science Foundation of Shanxi Province of PRC.

%\newpage

\begin{figure}
%\centering
%\epsfxsize=8.5cm
%\includegraphics[angle=-90,width=9.5cm]{fig1}
\caption{ Linear
 absorption $\alpha(\omega)$ plotted as a function of $(\omega-\omega_g)/\omega_{Bx}$,
  where $\omega_g$ is the frequency associated with the
  conduction-valence band gap.}

\end{figure}

\begin{figure}
%\centering \epsfxsize=8.5cm
%\includegraphics[angle=-90,width=9.5cm]{fig2}
\caption{The same as Fig.2, except that
$\omega_{By}:\omega_{Bx}=3:3=1:1$. Only one peak appears in every
fixed frequency interval $\omega_{bx}$.}
\end{figure}

\begin{figure}
%\centering \epsfxsize=8.5cm
%\includegraphics[angle=-90,width=9.5cm]{fig3}
\caption{ Spectrally-resolved degenerate squared four-wave mixing
signal $|P^{(3)}(\omega)|^2$ plotted as a function of
$(\omega-\omega_g)/\omega_{B,x}$.
  Three equidistant peaks appear within the fixed frequency interval
   $\omega_{Bx}$.}
\end{figure}

\begin{figure}
%\centering \epsfxsize=8.5cm
%\includegraphics[angle=0,width=8.5cm]{fig4}
\caption{Coherent time-evolution of a wavepacket, showing the
breathing behavior in the real-time domain. The oscillatory period
is $1$ ps, revealing the energy spacing in the frequency/energy
domain.}
\end{figure}


\begin{references}

\bibitem{mendez} E.E. Mendez and G. Bastard, Phys. Today, {\bf
50}, No.7, 30 (1993) and references therein.
\bibitem{meierl94}T. Meier, G. von Plessen, P. Thomas, S.W. Koch,
 Phys. Rev. Lett. {\bf 73}, 902 (1994); Phys. Rev. B {\bf 51}, 14490 (1995).
\bibitem{meierl95} T. Meier, F. Rossi, P. Thomas, and S.W. Koch,
 Phys. Rev. Lett. {\bf 75}, 2558 (1995).
\bibitem{axt} V.M. Axt, G. Bartels, and A. Stahl, Phys. Rev.
Lett. {\bf 76}, 2543 (1996).
\bibitem{whittaker} For Fano resonance in superlattices driven by dc field, see
D.M. Whitakker, Europhys. Lett. {\bf 31}, 55 (1995); Norbert
Linder, Phys. Rev. B {\bf 55}, 13664 (1997); C.P. Holfeld, F.
L\"oser, M. Sudzius, K. Leo, D.M. Whitakker, and K. K\"ohler,
Phys. Rev. Lett. {\bf 81}, 874 (1998); S. Glutsch, and F.
Bechstedt, Phys. Rev. B {\bf 60}, 16584 (1999); for dc-ac Fano
resonance case, see R.-B. Liu and B.-F. Zhu, J. Phys. Conden.
Matter, {\bf 12}, L741 (2000).
\bibitem{keay} B.J. Keay, S. Zeuner, S.J. Allen, K.D. Maranowski,
A.C. Gossard, U. Bhattacharya, and M.J.W. Rodwell, Phys. Rev.
Lett. {\bf 75}, 4102 (1995).
\bibitem{unter} K. Unterrainer, B.J. Keay, M.C. Wanke, S.J.
Allen, D. Leonard, G. Medeiros-Ribeiro, U. Bhattacharya, and
M.J.W. Rodwell, Phys. Rev. Lett. {\bf 76}, 2973 (1996).
\bibitem{dunlap} D.H. Dunlap, and V.M. Kenkre, Phys. Rev. B {\bf 34}, 3625 (1986).
\bibitem{jauho} A.P. Jauho, and K. Johnsen, Phys. Rev. Lett.
{\bf 76}, 4576 (1996).
\bibitem{loser} F. L\"oser, M.M. Dignam, Yu. A. Kosevich, K.
K\"ohler, and K. Leo, Phys. Rev. Lett. {\bf 85}, 4763 (2000).
\bibitem{zhao} X.-G. Zhao, R. Jahnke, and Q. Niu, Phys. Lett. A
{\bf 202}, 297 (1995); K.W. Madison, M.C. Fischer, and M.G.
Raizen, Phys. Rev. A {\bf 60}, R1767 (1999).
\bibitem{dignamprb99} M. Dignam, Phys. Rev. B {\bf 59}, 5770 (1999).
\bibitem{liuprb} R.-B. Liu and B.-F. Zhu, Phys. Rev. B {\bf 59},
5759 (1999).
\bibitem{hader} J. Hader, T. Meier, S.W. Koch, F. Rossi und N. Linder,
Phys. Rev. B. 55, 13799 (1997).
\bibitem{raizenpt} M. Raizen, C. Salomon, and Q. Niu  Phys. Today,
{\bf 50} No.7, 30 (1997)
 \bibitem{wilkinson} For experiment see, S.R. Wilkinson, C.F.
Bharucha, K.W. Madison; for theoretical interpretation, see Q.
Niu, and M.G. Raizen, Phys. Rev. Lett. {\bf 76}, 4512; Q. Niu,
X.-G. Zhao, G.A. Georgakis, and M.G. Raizen, {\it ibid}. {\bf 76},
4504.
\bibitem{dahan}M.B. Dahan, E. Peik, J, Reichel, Y.
Castin, and C. Salomon, Phys. Rev. Lett. {\bf 76}, 4508.
\bibitem{gluck} M. Gl\"uck, F. Keck, A.R. Kolovsky, and H.J. Korsch,
 Phys. Rev. Lett. {\bf 86}, 3116 (2001).
\bibitem{hofstader} D.R. Hofstader, Phys. Rev. B {\bf 14}, 2239
(1976).
\bibitem{reich} R.K. Reich, R.O. Grondin, and D.K. Ferry, Phys.
Rev. B {\bf 27}, 3483 (1983).
\bibitem{silin} A.P. Silin, Sov. Phys. USP. {\bf 28}, 972 (1985). (Usp. Fiz. Nauk, {\bf 147}, 485-521 (1985)).
\bibitem{lindberg} M. Lindberg, and S.W. Koch, Phys. Rev. B {\bf
38}, 3342 (1988); for a systematic review, see H. Haug, and S.W.
Koch, {\sl Quantum theory of the optical and electronic properties
of semiconductors}, 3rd. Ed. (World Scientific, Singapore), (1994)
and references
 therein.
\bibitem{davies} J.H. Davies, and J.W. Wilkins, Phys. Rev. B {\bf
38}, 1667 (1988).
\bibitem{yanjpcmprb} W. Yan, X.-G. Zhao, and H. Wang,  J. Phys.: Conden.
  Matter {\bf 10}, L11 (1998);   W. Yan, S.-Q. Bao, X.-G. Zhao, and J. Q. Liang,
 Phys. Rev. B {\bf 61}, 7269 (2000).
\bibitem{yanjpcm01} W. Yan, F. Claro, Z.Y. Zeng, and J.Q. Liang,
 J. Phys.: Conden. Matter {\bf 13}, 5103 (2001).
\bibitem{lindbergpra} M. Lindberg, R. Binder, S.W. Koch, Phys. Rev.
A {\bf 45}, 1865 (1992).
\bibitem{banyai} L. Banyai, D.B. Tran Thoai, E. Reitsamer, H.
Haug, D. Steinbach, U. Wehner, M. Wegner, T. Marschner, and W.
Stolz, Phys. Rev. Lett. {\bf 75}, 2188 (1995).
\bibitem{dignamprb94} For the wavepackets in the semiconductor superlattices driven by
 the one-dimensional electric field, see M. Dignam, J.E. Sipe, and J. Shah, Phys. Rev. B {\bf 49},
 10502 (1994).
\bibitem{hughes} S. Hughes, and D.S. Citrin, Phys. Rev. B {\bf
59}, R5288 (1999); S. Hughes, and D.S. Citrin, J. Opt. Soc. Am B
 {\bf 17}, 128 (2000).
\bibitem{note1} In order for the comparison of the Coherent
wavepackets at different time $t$, the longitudinal and transverse
depahsing rates $\Gamma_L$, $\Gamma_T$ were assumed to be
vanishing.
\bibitem{note2} We also calculated the case of $p:q=8:3$, where
eight clear-cut peaks appear in the frequency interval
$\omega_{Bx}$.

\end{references}
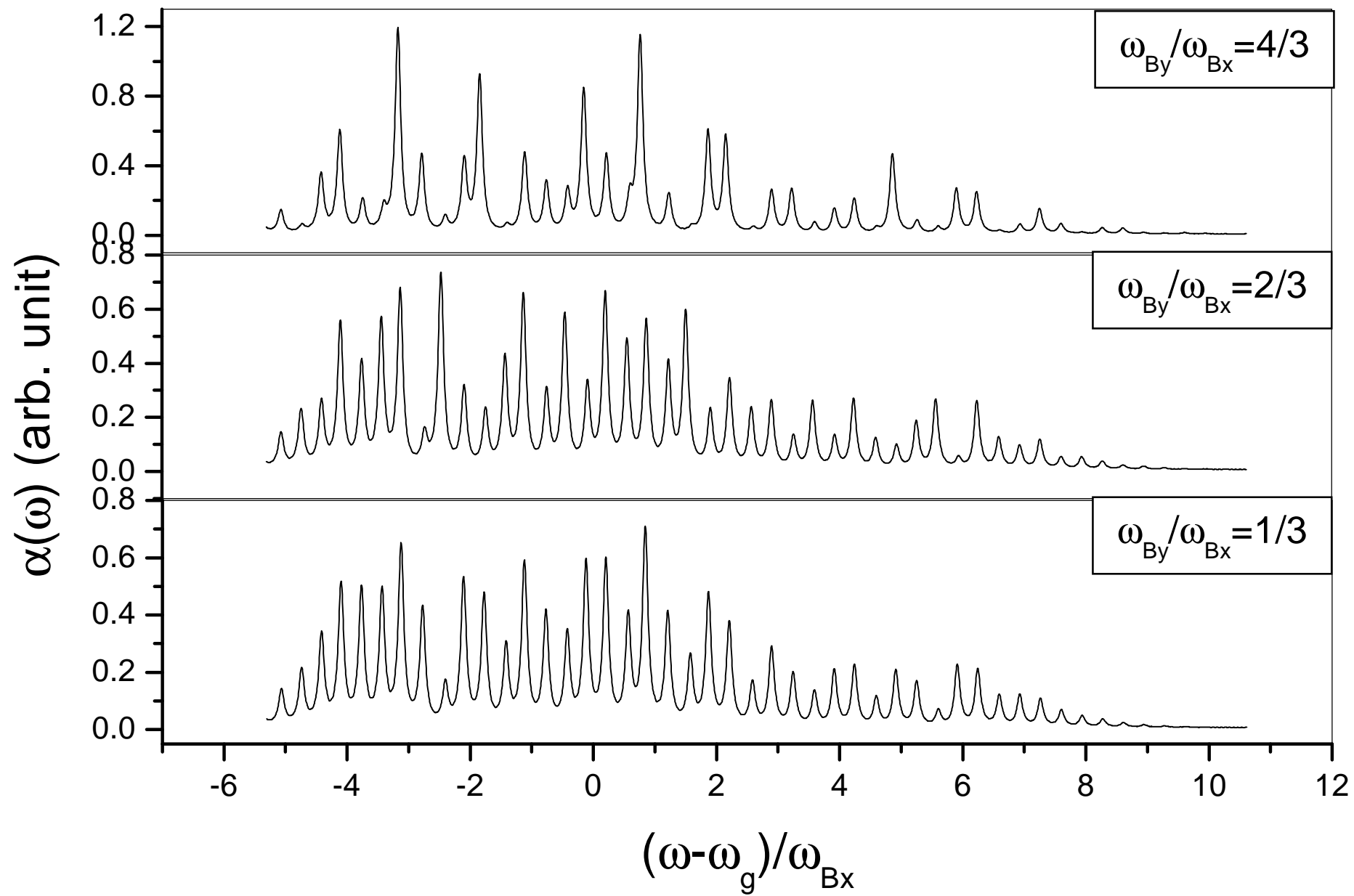
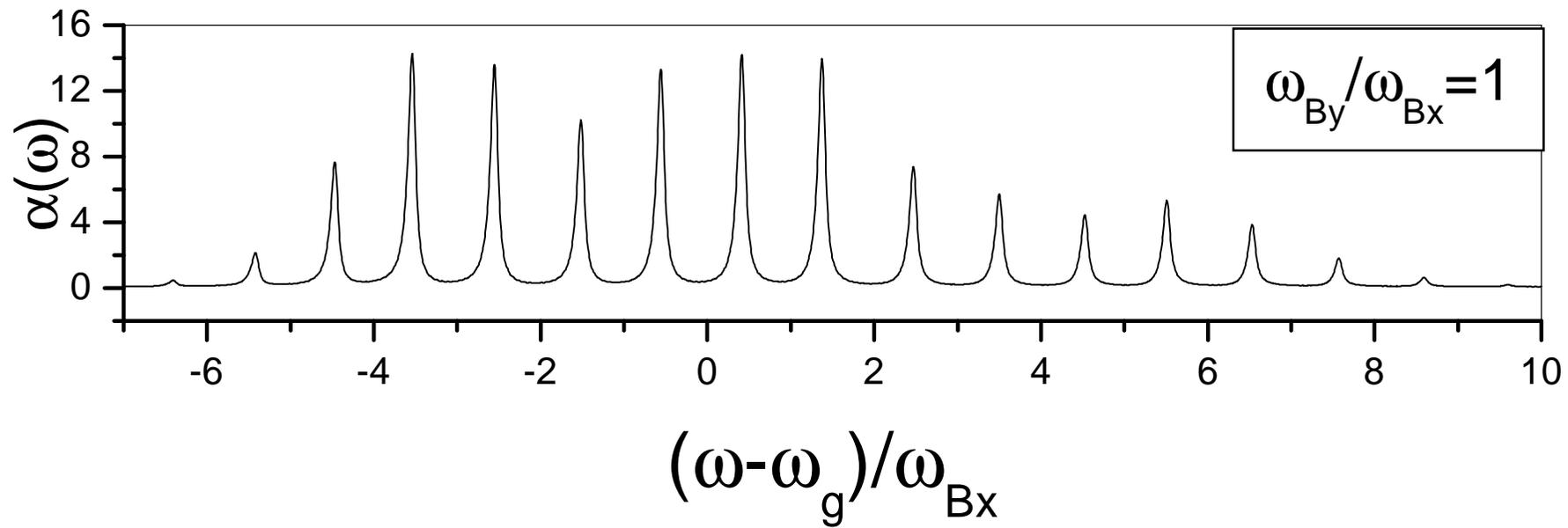
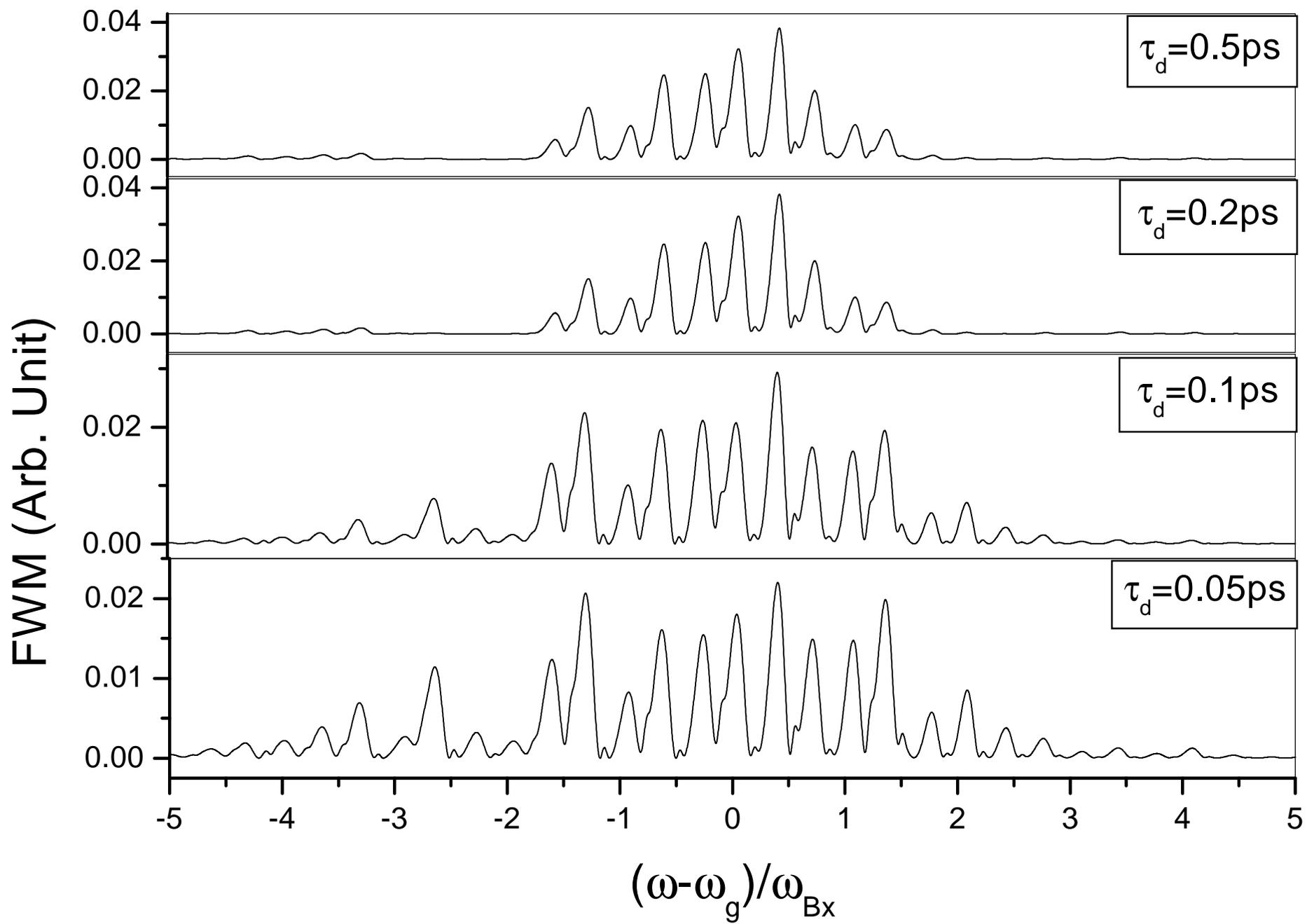
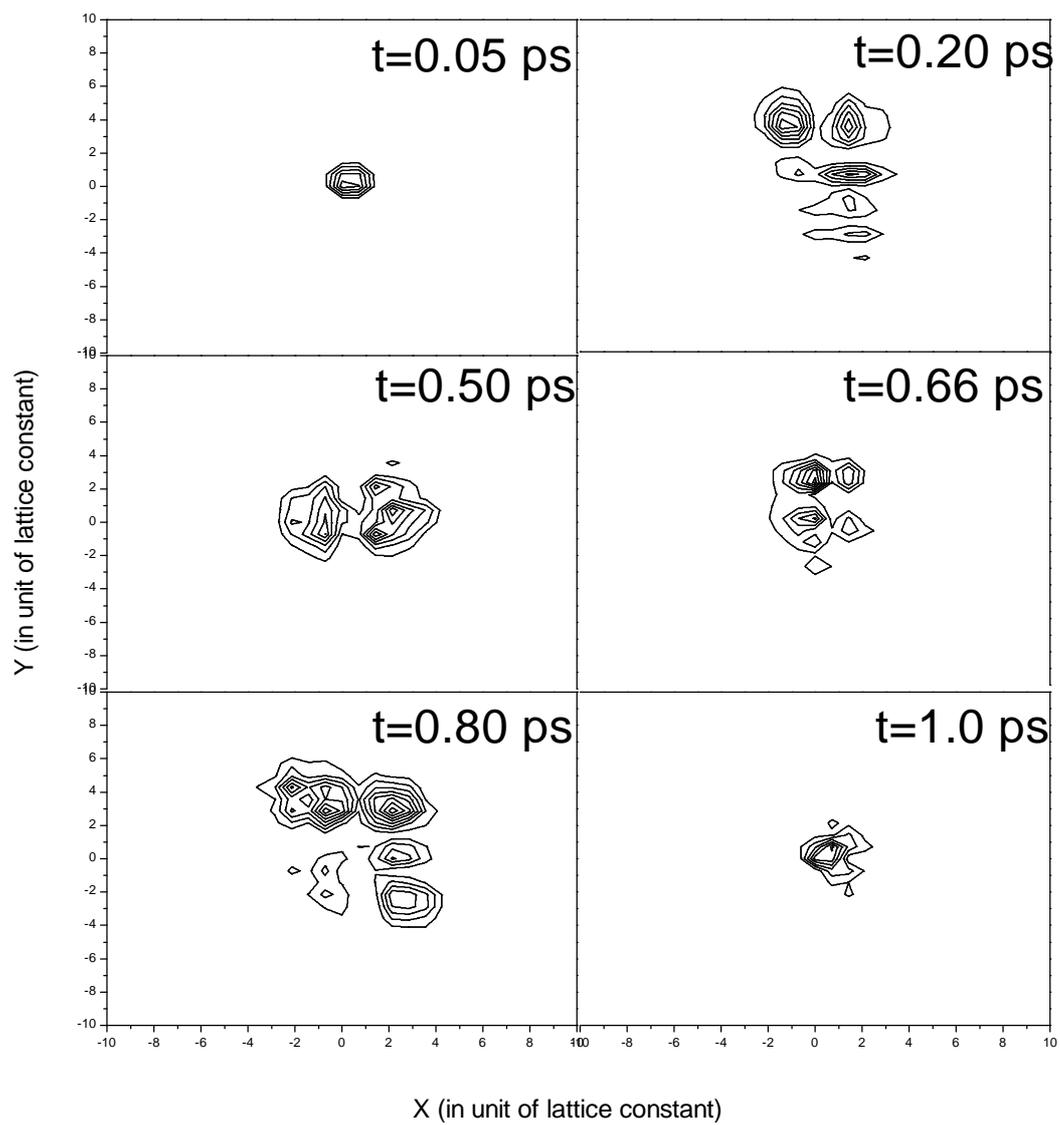
\end{document}